\documentclass{article}

\usepackage[english]{babel}

\usepackage[a4paper,top=2cm,bottom=2cm,left=3cm,right=3cm,marginparwidth=1.75cm]{geometry}

\usepackage{amsmath}
\usepackage{graphicx}
\usepackage[colorlinks=true, allcolors=blue]{hyperref}
\usepackage{comment}
\usepackage{placeins}
\usepackage[numbers,sort&compress]{natbib}
\usepackage{subcaption}
\usepackage{multirow}
\usepackage{amssymb}
\usepackage{siunitx}
\usepackage{booktabs}
\newcommand{\GeV}{\ensuremath{\text{Ge\kern -0.1em V}}}
\newcommand{\MeV}{\ensuremath{\text{Me\kern -0.1em V}}}
\newcommand{\keV}{\ensuremath{\text{ke\kern -0.1em V}}}

\title{Learning to bin: differentiable and Bayesian optimization for multi-dimensional discriminants in high-energy physics}

\author{Johannes Erdmann\footnote{Email: johannes.erdmann@physik.rwth-aachen.de}, Nitish Kumar Kasaraguppe\footnote{Email: nitish.kumar.kasaraguppe@rwth-aachen.de}, Florian Mausolf\footnote{Email: florian.mausolf@rwth-aachen.de}}

\date{\small
  RWTH Aachen University, III. Physikalisches Institut A, Aachen, Germany
  }

\begin{document}
\maketitle

\begin{abstract}

Categorizing events using discriminant observables is central to many high-energy physics analyses.
Yet, bin boundaries are often chosen manually.
A simple, popular choice in multi-classification tasks is to assign events according to the largest per-class score ("argmax") and to apply equidistant binning to the resulting one-dimensional discriminants.
We propose a binning optimization for signal significance directly in multi-dimensional discriminants.
We use a Gaussian Mixture Model (GMM) to define flexible regions in the score space, which can be interpreted either as bins or as analysis categories.
While this GMM-based strategy is applicable in both one and multiple dimensions, we also study a direct bin-boundary optimization in one dimension as a simpler alternative for binary discriminants.
On this binning model, we study two optimization strategies: a differentiable and a Bayesian optimization approach.
We study two toy setups: a binary classification and a three-class problem with two signals and backgrounds.
In the one-dimensional case, both approaches achieve similar gains in signal sensitivity compared to equidistant binning for a given number of bins, while in the multi-dimensional case the differentiable approach performs best.
We show that the GMM-based optimization can outperform argmax classification even after optimized binning is applied to the one-dimensional projections.
We further study the performance of our methods on the FAIR Universe $H\rightarrow\tau\tau$ dataset, where the GMM-based optimization gives the highest signal significance.
Both methods are released as lightweight Python plugins intended for straightforward integration into existing analyses.

\end{abstract}

\section{Introduction}
\label{sec:intro}
Modern collider experiments in high-energy physics (HEP) record enormous samples of particle collisions, enabling, for example, precision measurements of cross sections and searches for rare phenomena predicted by the standard model or its extensions.
In such analyses, statistical inference is often performed with binned likelihoods, where observed event yields in analysis categories or histogram bins are compared to a model in profile-likelihood fits.
Central to these fits is a discriminant that separates the targeted process(es) (``signal(s)'') from backgrounds.
Depending on the analysis, this discriminant may be a simple kinematic observable (e.g., a reconstructed particle’s transverse momentum) or the output of a machine-learning classifier. 
Such classifiers are now standard in HEP, in both binary and multi-class forms, producing one- or multi-dimensional output distributions, respectively.

The choice of binning directly impacts the sensitivity of the analysis:
By grouping regions of the discriminant space into bins, information is discarded, which reduces the sensitivity to the signal(s) but ensures that signal and background predictions in the different bins are reliable.
In practice, bin boundaries are often chosen manually.
Common choices include equidistant or quantile-based binning in one dimension.
Events with multi-dimensional classifier scores are often categorized according to the class with the maximum score ("argmax") and then binned in the one-dimensional projections.
While straightforward, these choices are not guaranteed to be optimal for the analysis figure of merit, especially for multi-dimensional outputs where information can be lost in the projection.

Only a limited number of approaches have so far directly addressed this analysis-optimization aspect.
Adaptive “Bayesian Blocks” histogramming~\cite{Pollack:2017srh} defines variable-width bins by maximizing a Poisson likelihood for piecewise-constant rates within these bins, penalized for adding more partitions. It hence targets density-estimation quality in one-dimensional distributions rather than an analysis figure of merit.
Several other approaches focus on \emph{analysis-aware} optimization of categorization.
ThickBrick~\cite{Matchev:2019lfq} finds binning thresholds that explicitly optimize common significance proxies, possibly depending on another event variable for decorrelation, in an iterative $k$-means-like procedure.
The INFERNO method~\cite{DeCastro:2018psv} learns summary statistics by passing events through a multi-output neural network where each output component is regarded as a bin, applying a softmax to obtain differentiability during training, and optimizing an inference-aware loss derived from the Asimov Fisher information of a Poissonian likelihood with nuisance parameters.
Wunsch et al.~\cite{Wunsch:2020iuh} apply a related inference-aware objective, where the binning in a one-dimensional summary variable is fixed a priori and the histogramming step is made differentiable by replacing hard bin boundaries by narrow Gaussian kernels.
The end-to-end differentiable analysis pipeline \textit{neos}~\cite{Simpson:2022suz} replaces non-differentiable histogramming by a binned kernel density estimator (KDE), selection cuts are replaced by sigmoid functions, and the full profile likelihood is differentiated, so that the network parameters can be optimized for expected sensitivity. 
In this KDE-based approach, however, the bin edges themselves are kept fixed.
A complementary idea is to construct categories directly in the multi-class score space.
In Ref.~\cite{Diekmann:2024oph}, the classifier output is clustered with a $k$-means algorithm and each cluster defines a category in the inference, preserving correlations but following the event-density structure of the score space rather than a metric that is directly derived from the analysis sensitivity.

In this work, we address the binning problem directly with respect to a chosen analysis objective, using the expected signal significance as one illustrative example in the benchmark studies presented here.
We seek bin boundaries for a trained classifier discriminant that maximize the expected signal significance without requiring one-dimensional projections or rectangular regions in the multi-dimensional space.
Our approach is built on a Gaussian Mixture Model (GMM), where each bin is represented by a Gaussian component and events are assigned to the component with the highest probability. 
We optimize the position (mean) and shape (covariance) of each Gaussian component to maximize the objective.

We consider two complementary optimization strategies and make them available to the community: a differentiable programming approach, and a Bayesian optimization approach using a tree-structured Parzen estimator~\cite{NIPS2011_86e8f7ab}.
In the following, we refer to the differentiable approach as GATO (Gradient-based cATegorization Optimizer), which is based on automatic differentiation in TensorFlow~\cite{abadi2016tensorflowlargescalemachinelearning} and label the Bayesian approach BOBR (Bayesian Optimization of Bin boundaRies).
We provide lightweight plugins, intended for straightforward integration into analysis workflows with one- or multi-dimensional discriminants, as the Python packages \texttt{gato-hep} and \texttt{bobr-hep} (available on PyPI and GitHub\footnote{bobr-hep: \href{https://github.com/nitishkumarkv/bobr-hep}{https://github.com/nitishkumarkv/bobr-hep}}, respectively).
Complementary to end-to-end optimization approaches such as INFERNO and \textit{neos}, GATO and BOBR can be applied post-training to fixed classifier output scores in existing analysis workflows, while GATO can also be embedded in larger differentiable workflows.
It may be used as one piece of the joint optimization of a classifier and its binning.
Or it may be a piece in a fully differentiable analysis pipeline made of several differentiable building blocks, such as differentiable selections~\cite{Hance:2025haf}, classifier training, binning, and a differentiable statistical analysis~\cite{Feickert:2022lzh, evermore}.

\section{Dataset}
\label{sec:Dataset}

To illustrate and benchmark the binning–optimization methods proposed in this study, we construct toy datasets that represent typical discriminant shapes encountered in HEP analyses.
In all cases, events are described by a three-dimensional feature vector serving as a proxy for three discriminating variables.
We include one or two signal processes and five background processes.
For each physics process, events are generated by sampling these three-dimensional feature vectors from a multivariate Gaussian distribution. Events are assigned weights according to dummy cross sections, reproducing imbalanced yields per process, as in typical analyses.

To avoid unrealistically sharp process separation, we introduce a small 
multiplicative perturbation to the sampled observables.
On top of these features, we construct toy discriminants that mimic the output of multivariate classifiers.
We consider two illustrative discriminants constructed from the same underlying three-dimensional features: a one-dimensional binary classification score and a three-class softmax-like output.

For the one-dimensional example, we build the likelihood ratio for a single signal against the combined background and normalize it to the interval $[0,1]$. 
For the three-dimensional example, we construct softmax-like scores for the two signals and the combined background, yielding a three-class output whose components sum up to one.
In contrast to a single realistic example, the toy setup allows us to vary the properties of the discriminant and study the performance of the optimization methods in different scenarios.
For this purpose, we compare two scenarios obtained by varying the Gaussian mean vectors of the generated processes, leading to different signal and background distributions in the classifier-output space.
The precise parameter values and analytic formulae underlying both toy setups are documented in Appendix~\ref{app:dataset}.
Although systematic effects are not included in the toy datasets, their potential integration through nuisance-aware optimization objectives is discussed in Section 3.

\section{Binning model and optimization approaches}
\label{sec:binning_model}

The aim of the binning model choice is to provide a flexible, yet mathematically simple parameterization that can represent bin boundaries of non-trivial shapes in one- or multi-dimensional discriminant spaces.
To this end, we describe the discrete assignment of events into \(K\) bins by a GMM with \(K\) components.
The dimensionality $d$ of the Gaussian components is chosen to match the dimensionality of the underlying discriminant.

For an event with discriminant value $x\in\mathbb{R}^d$, each Gaussian component with mean $\boldsymbol{\mu}_k$ and covariance matrix $\Sigma_k$ defines the likelihood
\begin{equation}
  \ell_k(x)
  = \mathcal{N}\!\bigl(x \mid \boldsymbol{\mu}_k,\, \Sigma_k\bigr).
\end{equation}
Together with a mixture weight $\pi_k$ of the component, we define the unnormalized logarithmic score
\begin{equation}
  s_k(x) = \log \ell_k(x) + \log \pi_k.
\end{equation}
The binning is performed such that an event is assigned to the component with the largest score $s_k(x)$.

The mixture weights are implemented starting from unconstrained trainable parameters which are normalized with the log-softmax function to keep them strictly positive.
The means of the Gaussian components are constrained so that they do not exceed the discriminant domain.
For discriminants living on a simplex, as in our three-class example, we omit one coordinate of the $d$-dimensional score vector (since the components sum to unity) and construct the means in the remaining $d-1$ directions.
We then append a fixed zero as the $d$-th entry and apply the softmax, ensuring the learned means reside on the same simplex.

To guarantee positive definiteness of the covariance matrices while keeping the parameterization unconstrained, each covariance matrix is represented via a Cholesky factorization,
\begin{equation}
  \Sigma_k = L_k\,L_k^{T},
\end{equation}
with $L_k$ being a lower-triangular matrix.
For the Cholesky factors of the covariances, we start with unconstrained trainable tensors $L^{\rm raw}_k$.
We set the parts
\begin{equation}  
  O_k = \kappa\cdot \mathrm{lower}(L^{\rm raw}_k), 
  \quad
  D_k = \exp\bigl(\mathrm{diag}(L^{\rm raw}_k)\bigr),
\end{equation}
as off-diagonal and diagonal elements of the Cholesky factor $L_k = O_k + D_k$, respectively. Here, $\mathrm{lower}(\cdot)$ zeroes the diagonal and all entries above it.
The element-wise exponential ensures strictly positive variances, and the hyperparameter $\kappa$ damps the evolution of off-diagonal elements (i.e., correlations), which we found to improve training stability.
In all experiments presented here, we use $\kappa = 0.1$.

This formulation allows the Gaussian components to take on general elliptical shapes, including rotations and correlations between dimensions.
Importantly, the resulting bins are not constrained to ellipsoidal shapes: since events are assigned to the component with the highest probability density, the boundaries between components arise from comparing multiple Gaussian densities. These boundaries can therefore be highly non-linear, even when each component itself has a simple elliptical form.
The GMM-based representation thus offers high flexibility to approximate complex bin shapes while keeping the number of trainable parameters manageable for efficient optimization.

While in principle, any continuous function of the binned signal and background yields, $S$ and $B$, could be used as figure of merit, we use the Asimov significance~\cite{Cowan:2010js} here.
For the $k$-th bin, it is given by 
\begin{equation}      
   Z_k \;=\;\sqrt{\,2\Bigl[(S_k+B_k)\,\ln\!\bigl(1+\tfrac{S_k}{B_k}\bigr)-S_k\Bigr]\,}\,,
\end{equation}
and added in quadrature over all bins.
In realistic analyses, bins with extremely low background yields or excessively large statistical uncertainties can be problematic.
To guide the optimization away from such configurations, we introduce two optional penalty terms.

First, to discourage bins with effectively zero background yield, we add a
quadratic penalty when $B_k$ falls below a user-defined minimum threshold $B_{\min}$:
\begin{equation}
  P_{\rm low}(B)
  = \sum_{k=1}^K \bigl[\max(0,\,B_{\min} - B_k)\bigr]^2.
\end{equation}

Second, to suppress bins with a large relative statistical uncertainty in the background prediction, we penalize bins whose uncertainty exceeds a limit $r$.
For a given bin $k$, the expected background yield is obtained as a weighted sum of events,
\begin{equation}
  B_k = \sum_{i} w_i ,
\end{equation}
where $w_i$ denotes the weight of event $i$ assigned to that bin.
Assuming independent Poisson fluctuations of the underlying event counts, the variance of a weighted sum is given by the sum of squared weights,
\begin{equation}
  \mathrm{Var}(B_k) = \sum_{i} w_i^2 .
\end{equation}
The corresponding relative statistical uncertainty in the background prediction in bin $k$ is therefore
\begin{equation}
  \sigma_{\mathrm{rel},k}
  = \frac{\sqrt{\sum_{i} w_i^2}}{B_k} .
\end{equation}
This leads to the penalty definition
\begin{equation}
  P_{\rm unc}(B)
  = \sum_{k=1}^K \bigl[\max(0,\,\sigma_{\mathrm{rel},k} - r)\bigr]^2.
\end{equation}
These terms can be weighted and combined with the main objective.

Although the benchmark studies in this work only target the signal significance with respect to statistical uncertainties, the optimization framework can be extended to nuisance-aware objectives.
Systematic effects could be incorporated either by augmenting the objective with additional regularization terms that disfavor binning choices strongly affected by such variations, or by replacing the nominal significance with a differentiable inference-aware objective that accounts for nuisance parameters. 
Such a target could, for example, be constructed using the Fisher information of a Poisson likelihood or a differentiable profile-likelihood-based objective, following the general direction of inference-aware optimization approaches~\cite{DeCastro:2018psv,Simpson:2022suz,Wunsch:2020iuh}.
In this way, optimization would favor not only binning schemes that are sensitive to the nominal model, but also those that remain robust against relevant systematic variations.

In the following sections, we describe the two complementary strategies, GATO and BOBR, that we use to optimize the GMM parameters with respect to the objective.

\subsection{GATO: gradient-descent based approach}
\label{sec:gato}

The key requirement for the GATO approach is a differentiable approximation of the bin assignment.
We obtain this by replacing the hard assignment of an event to a specific bin with a temperature-scaled softmax over the scores $s_k(x)$.
For an event with discriminant value $x$, the fractional assignment to bin $k$ is given by
\begin{equation}  
  \gamma_k(x) = \operatorname{softmax}_k\!\left(\frac{s_k(x)}{T}\right),
\end{equation}
which approaches step-like bin boundaries in the limit $T \to 0$ and yields smoother gradients for larger $T$.
Expected bin yields are calculated by summing the event weights, additionally weighted by these fractional bin memberships.

To avoid poor local minima and encourage exploration early on, we start with $T=T_0=1$ and gradually decay to $T_f\ll1$.
We use a cosine annealing schedule to update the temperature after each epoch and apply this scheme also to the learning rate.
We use the RMSprop optimizer~\cite{Hinton2012RMSProp} to update the trainable parameters once per epoch.

For one-dimensional discriminants, we also provide a simpler alternative in \texttt{gato-hep}, where bin boundaries are represented directly by trainable cut values.
This follows the standard differentiable programming approach of replacing hard boundaries with smooth sigmoid approximations and provides a validation for the more general GMM-based strategy in the one-dimensional case.
Each boundary is replaced by a sigmoid function that acts as a differentiable approximation to a step function, so that events receive fractional weights on either side of the boundary.
The cuts are constrained to remain strictly ordered by implementing them as positive cumulative increments, which prevents them from crossing.
For multiple bins, the sigmoid transitions are combined to produce a set of smooth, non-negative bin weights that sum to one.
This construction can also be extended to rectangular selections in several variables by forming one-dimensional sigmoid weights for each variable and taking their product to obtain the smooth weight of the multi-dimensional region.

\subsection{BOBR: Bayesian optimization approach}

As a complementary strategy to the gradient-based GATO method, which relies on differentiable approximations and gradient descent, BOBR applies Bayesian optimization~\cite{10.1115/1.3653121, Mockus1974OnBM, Zhilinskas1975} to determine the bin boundaries by maximizing the objective chosen through probabilistic modeling.
This results in a sequential optimization procedure that balances exploration of the binning parameter space with exploitation of regions that yield higher objective values, a characteristic feature of Bayesian optimization.
While such methods are most commonly used for hyperparameter optimization in machine-learning applications, we show that they can also be applied to bin-boundary optimization.

In this work, BOBR is implemented using a Tree-structured Parzen Estimator (TPE) algorithm~\cite{NIPS2011_86e8f7ab}, as provided by the \texttt{Optuna}~\cite{10.1145/3292500.3330701} framework. 
The optimization objective is defined as the asymptotic significance introduced earlier, with the penalty terms subtracted from the objective to regularize the binning configuration.
To improve the robustness of the optimization in situations where the objective remains non-positive over several iterations, we employ an adaptive strategy within the TPE procedure.
After a specified, user-defined number of trials without improvement, the TPE parameter \(\beta\)~\cite{watanabe2023tree}, which controls the partitioning of objective values into better- and worse-performing regions of the parameter space, is halved and the optimization is restarted.
This strategy enables the optimization to move beyond sub-optimal configurations and explore a broader region of the binning parameter space, while maintaining the regularizing effect of the penalty terms.

Since the BOBR approach does not require a differentiable approximation of the bin edges, the one-dimensional case allows for a direct optimization of the bin edges.
For a binning with \(n\) bins defined on the discriminant interval \([0,1]\), the \(n-1\) internal bin edges are treated as optimization parameters whose ordering is enforced by construction.
In particular, the first edge is sampled in the range
\[
e_1 \in [0,1],
\]
and the remaining edges are sampled sequentially in the range
\[
e_i \in [e_{i-1}, 1], \quad i = 2, \ldots, n-1\,.
\]

For the multi-dimensional optimization problem, we employ the TPE sampler in \texttt{Optuna} with the \texttt{multivariate=True}~\cite{watanabe2023tree} option enabled, such that a multivariate kernel density estimate is used to model the joint GMM parameter distribution, rather than treating parameters independently. 
This formulation allows correlated variations of component means, covariances, and mixture weights to be sampled consistently when proposing candidate binning configurations. 
The number of startup trials is set to $100$ for all multi-dimensional optimization runs to obtain a reliable initial estimate of the high-dimensional GMM parameter space, enabling more effective sampling of subsequent trials in high-performing regions of the parameter space.

For both, the one-dimensional and multi-dimensional optimization studies, trials are partitioned into better- and worse-performing sets according to a square-root-based quantile rule applied to the objective values.
In this approach, the percentile threshold separating the two sets is scaled by the
inverse square root of the number of completed trials, which promotes exploration~\cite{watanabe2023tree}.

\section{Results}
\label{sec:results}

In the following section, we present the performance of the two complementary binning and optimization strategies, GATO and BOBR, as applied to the toy discriminants introduced in Section~\ref{sec:Dataset} and to the FAIR Universe $H\rightarrow\tau\tau$ dataset, a public high-energy physics benchmark dataset.

Unless stated otherwise, all results are derived using a default configuration for the underlying toy datasets. 
For each study, we generate \(100,000\) events  for every signal process and a total of \(1,000,000\) background events. 
Additionally, each optimization applies penalties to bins that have fewer than one expected background event or a relative background uncertainty exceeding \(50\%\). This approach prevents the formation of empty or statistically unstable bins.
Unless stated otherwise, the penalty strengths are set to \(\lambda = 10\) for BOBR and to \(\lambda = 0.2\) for the GATO-based methods, for both the minimum background yield and relative uncertainty constraints.

\subsection{Binary classification example}
\label{sec:1d_results}

We first demonstrate the behavior of the optimization strategies using the one-dimensional toy discriminant in the first scenario.
For each choice of 5, 10, and 20 bins, we evaluate every method in 20 independent optimization runs, each initiated from a different random starting configuration.

\begin{figure}[t]
    \begin{subfigure}[b]{0.45\linewidth}
        \centering
        \includegraphics[width=\linewidth]{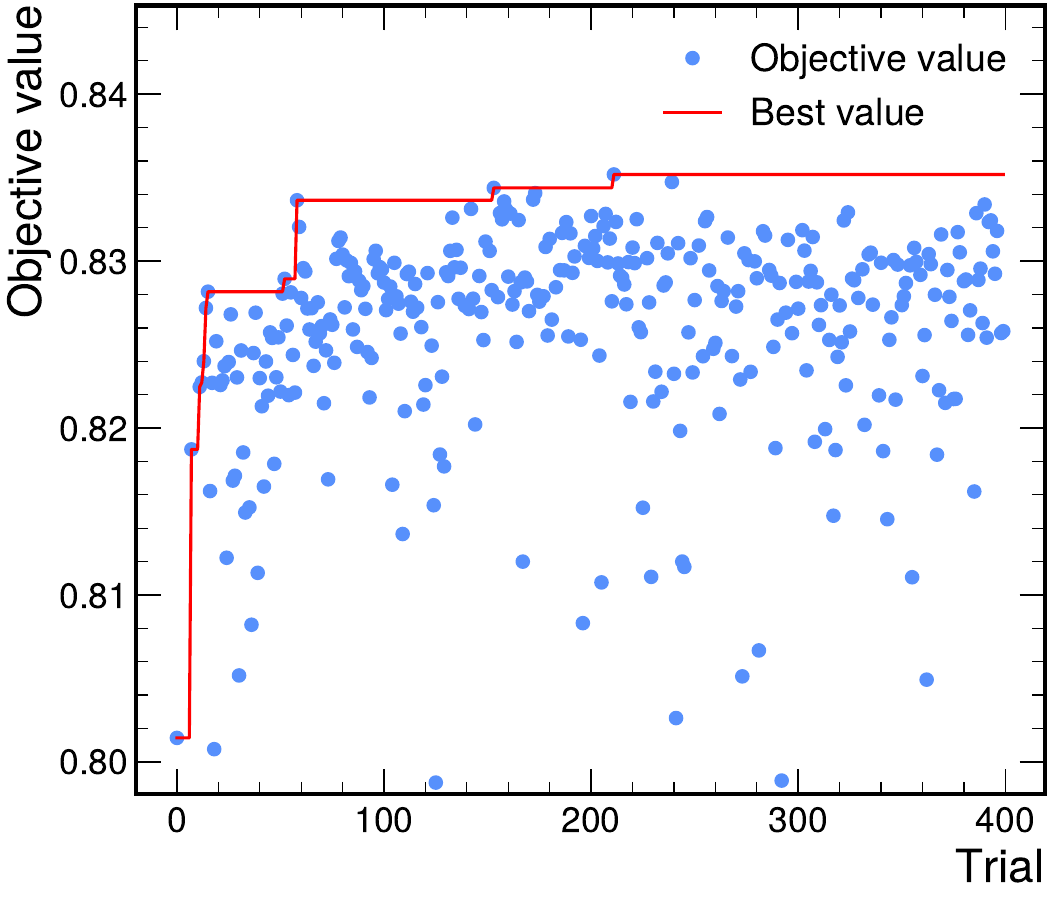}
        \caption{BOBR}
        \label{fig:optimization_history_plot_bobr_1d}
    \end{subfigure}
    \hfill
    \begin{subfigure}[b]{0.45\linewidth}
        \centering
        \includegraphics[width=\linewidth]{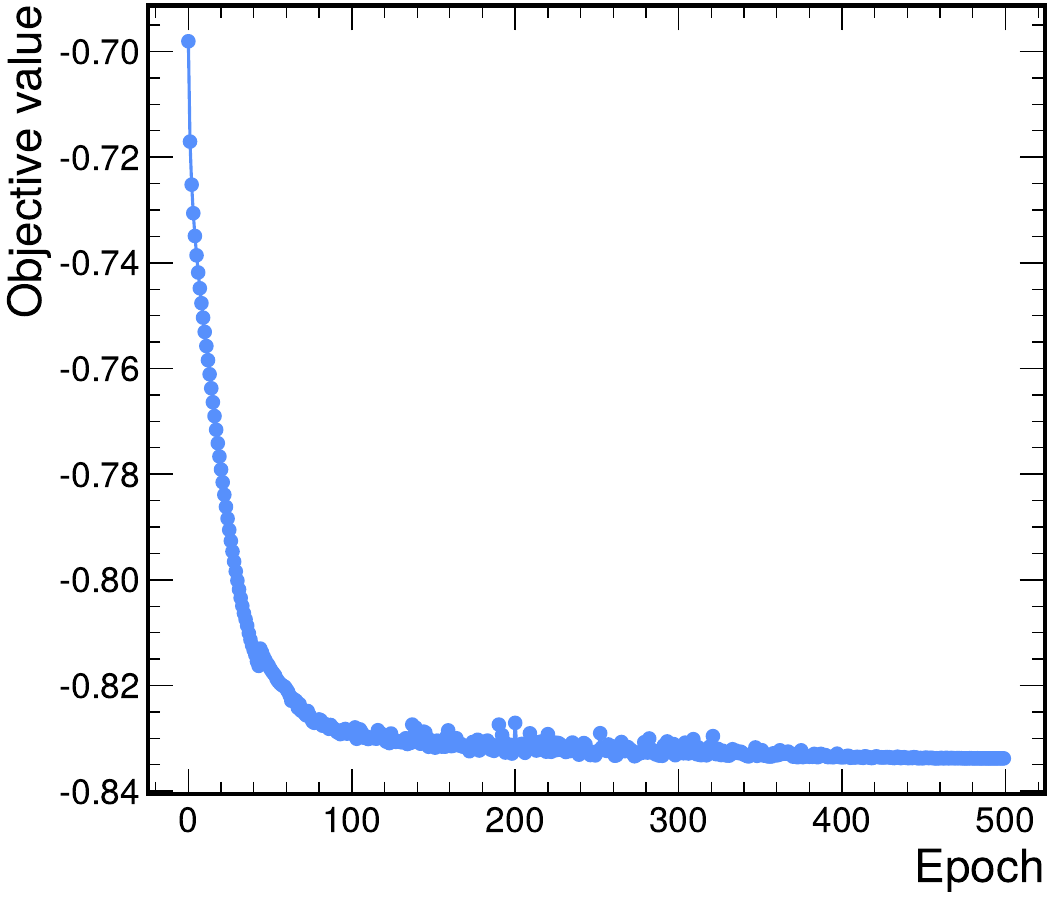}
        \caption{GATO sigmoid}
        \label{fig:history_loss_5bins_gato_1d_sigmoid}
    \end{subfigure}

    \vspace{0.5cm}
    \centering
    \begin{subfigure}[b]{0.45\linewidth}
        \centering
        \includegraphics[width=\linewidth]{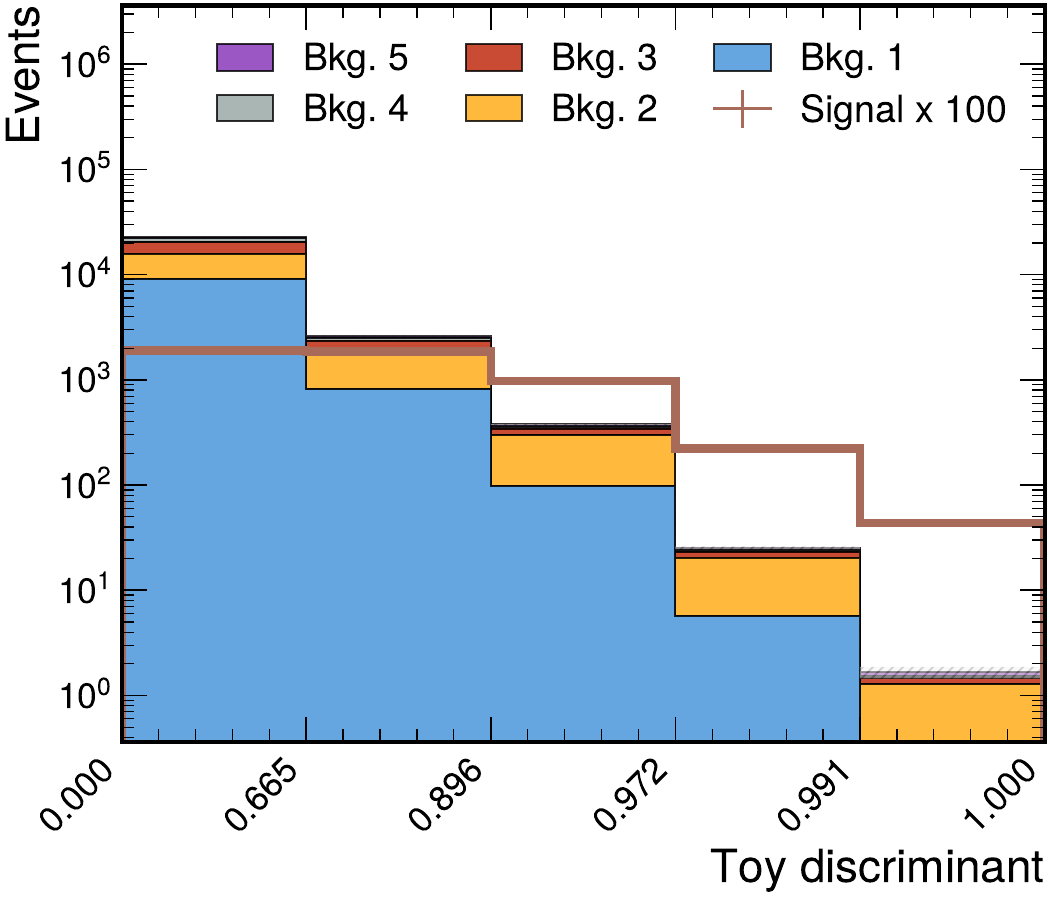}
        \caption{BOBR}
        \label{fig:plot_log_bobr_1d}
    \end{subfigure}
    \hfill
    \begin{subfigure}[b]{0.45\linewidth}
\        \centering
        \includegraphics[width=\linewidth]{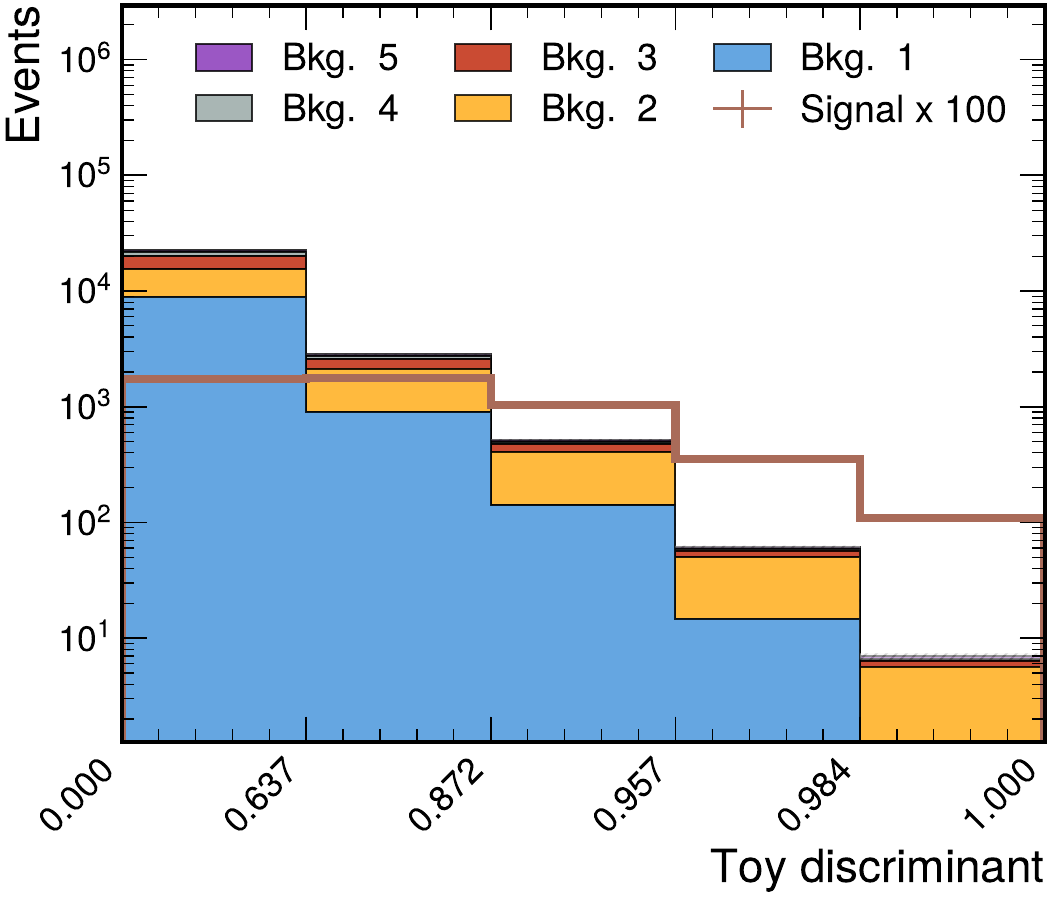}
        \caption{GATO sigmoid}
        \label{fig:NN_output_distribution_optimized_5bins_log_gato_1d_sigmoid}
    \end{subfigure}
    \vspace{0.5cm}
    \caption{Top row: Optimization histories from a representative trial for BOBR and GATO with a sigmoid-based parameterization.
Bottom row: Five-bin distributions for the same trials.}
    \label{fig:1D_results}
\end{figure}

The top row of Fig.~\ref{fig:1D_results} illustrates the optimization history for a representative trial for BOBR and GATO with a sigmoid binning strategy.
The BOBR objective demonstrates significant fluctuations between successive evaluations, as different bin-boundary configurations are explored. In contrast, GATO-based optimization exhibits a smooth and steady convergence toward the minimum.
The bottom row of Fig.~\ref{fig:1D_results} shows the corresponding five-bin distributions.
All methods exhibit comparable boundary placements, including the GATO with GMM parameterization, which is not shown.

\begin{figure}[h!]
    \centering
    \includegraphics[width=0.45\linewidth]{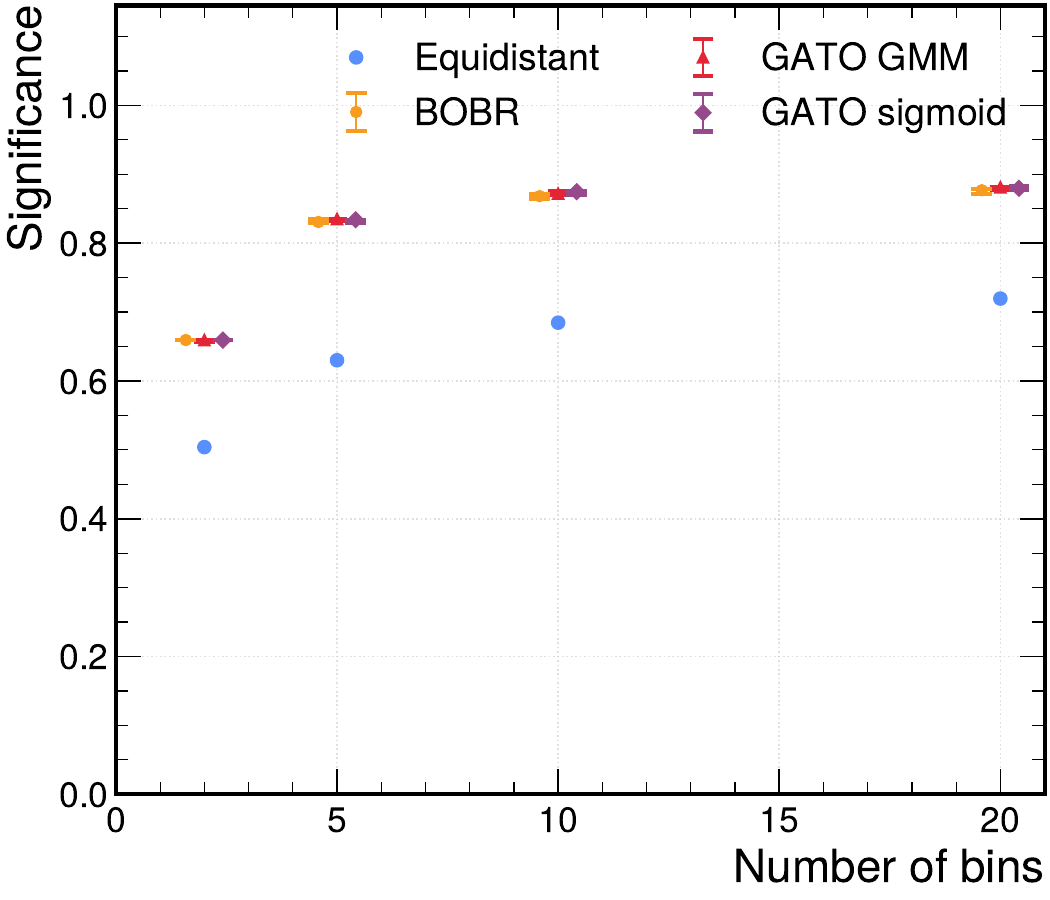}
    \caption{Comparison of the significance obtained in the one-dimensional setup.
    Error bars are obtained from independent optimization runs, marking the 16th to 84th percentiles.
    The markers for the BOBR and GATO significances are shifted horizontally around the tested bin counts to enhance the visibility.
    }
    \label{fig:1d_comp}
\end{figure}

The median Asimov significance of different runs is illustrated in Fig.~\ref{fig:1d_comp}, with error bars representing the interval between the 16th and 84th percentiles.
As a baseline, we also consider an equidistant binning of the discriminant output in the interval \((0,1)\). 
The three optimized strategies -- BOBR, GATO with GMM parameterization, and GATO with sigmoid-based parameterization -- yield very similar significances across all bin counts, mostly agreeing within the indicated error bars.
BOBR provides a derivative-free alternative for cases where the objective is not available in a differentiable form.
As mentioned in Section~\ref{sec:gato}, the sigmoid-based approach provides a validation for the more general GMM-based strategy in the one-dimensional case.
By contrast, the equidistant binning yields lower significances, as expected.
Achieving a performance comparable to the results of the optimization strategies proposed in this study would require a substantially larger number of bins. 
This illustrates the advantage of adapting the bin boundaries to the structure of the underlying discriminant.

\subsection{Three-class example}
\label{sec:multi_dim_results}

As discussed previously for the GMM-based binning strategy, the three-class discriminant lies on a simplex, allowing us to omit the background component and build the Gaussian mixtures in the remaining two signal-score components. 
Throughout this section, we define the figure of merit as the geometric mean of the two signal significances, in order to achieve a class-balanced categorization. 

As a baseline that is motivated by common practice in HEP analyses, we consider an argmax categorization of the discriminant, in which each event is assigned to the class corresponding to the highest output score. 
In this setup, only the two signal categories are retained, and an equidistant binning is applied separately for each of them; the background argmax region is not used in this comparison.
For consistency with the total bin counts used in the optimized methods, a setup with \(N\) bins uses \(N/2\) bins in each of the two signal categories.

In addition to this baseline, we include two further methods for comparison. 
The first replaces the equidistant binning with a one-dimensional BOBR optimization performed independently in each signal argmax category. 
As observed in Section~\ref{sec:1d_results}, the one-dimensional variants of GATO perform similarly to BOBR, and we would therefore expect a comparable behavior if a GATO-based one-dimensional binning was used instead. 
Finally, we include $k$-means clustering on the two-dimensional signal-score plane, following Ref.~\cite{Diekmann:2024oph}, where clusters of the classifier output are used directly as bins, providing an additional point of reference.

We compare the performance of the GMM-based binning for both GATO and BOBR, together with the other approaches, in two representative scenarios of the three-dimensional toy discriminant.
The two scenarios are chosen to illustrate different configurations in the multi-dimensional score space. In the first scenario, the two signals and the background overlap significantly in the signal-score plane. In the second scenario, the signal-enriched regions are more localized.

\begin{figure}[p]
    \centering
    \begin{subfigure}[b]{0.45\linewidth}
        \centering
        \includegraphics[width=\linewidth]{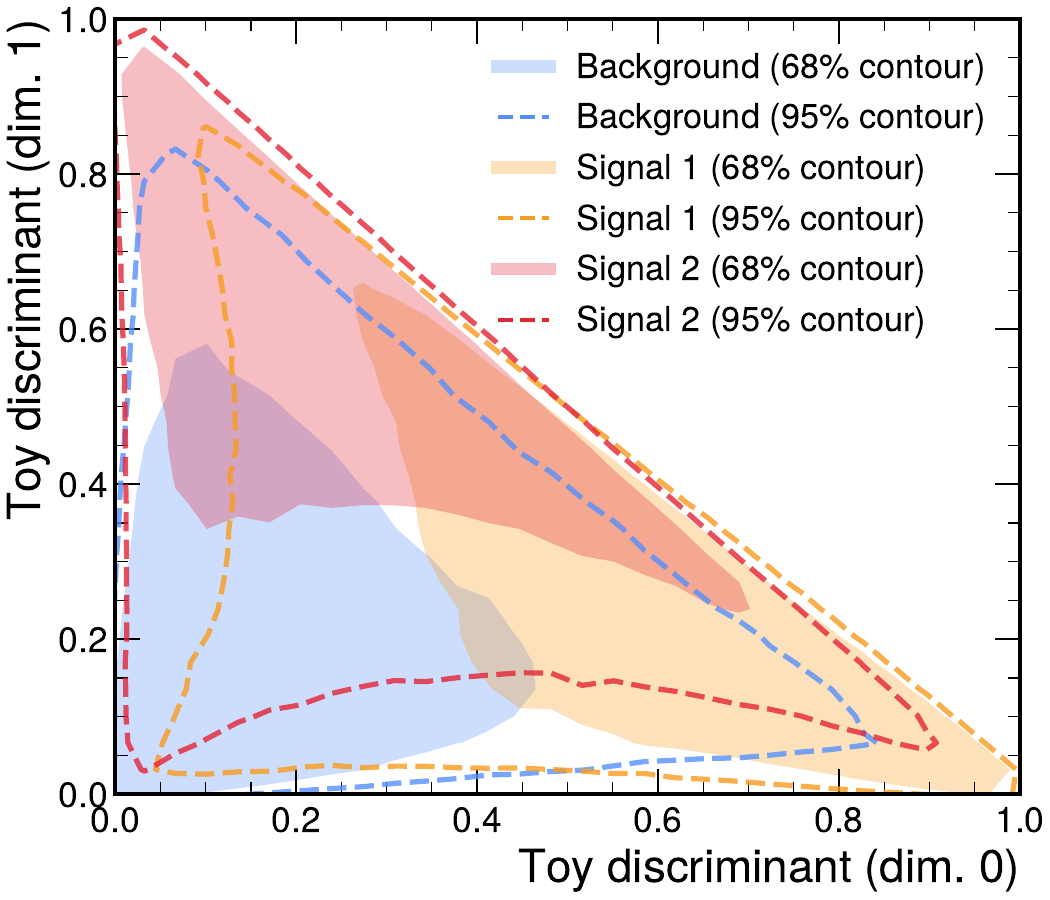}
        \caption{First scenario}
        \label{fig:contour_prob_toy_data_3class_filled_scene_1}
    \end{subfigure}
    \hfill
    \begin{subfigure}[b]{0.45\linewidth}
        \centering
        \includegraphics[width=\linewidth]{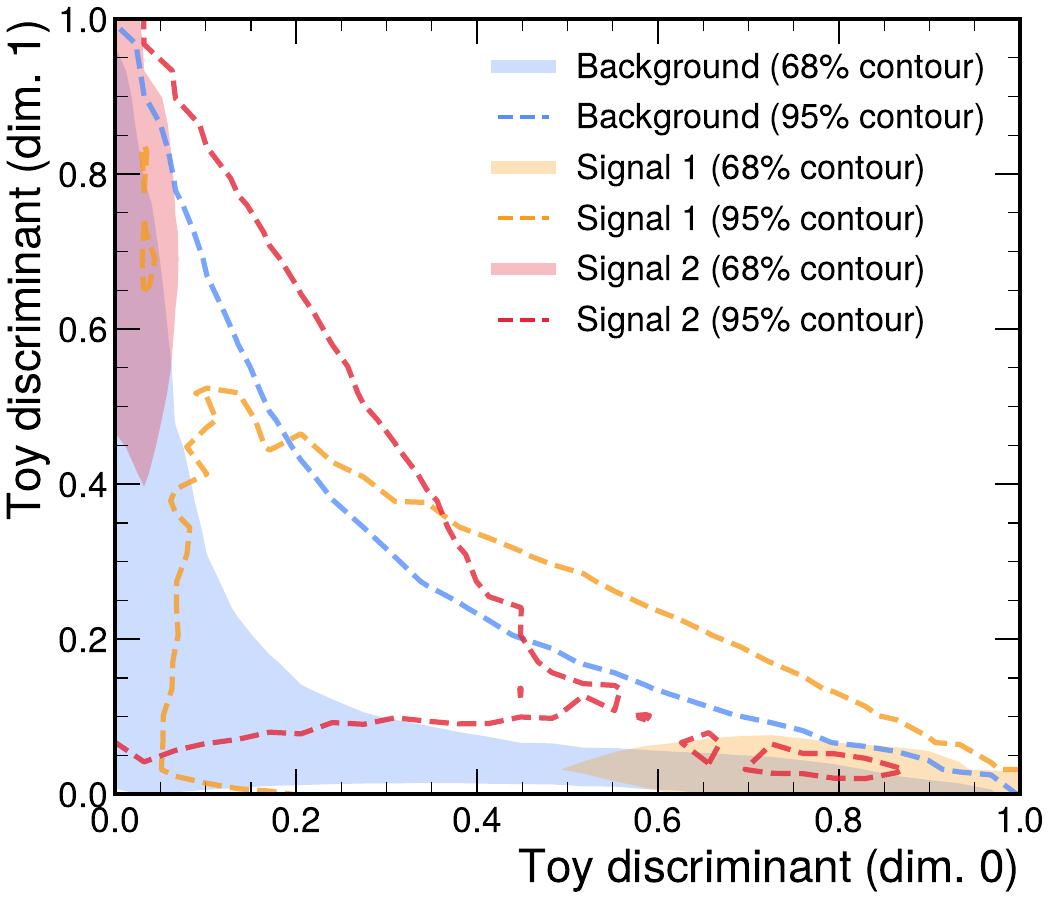}
        \caption{Second scenario}
        \label{fig:contour_prob_toy_data_3class_filled_scene_2}
    \end{subfigure}
    \vspace{0.5cm}
    \begin{subfigure}[b]{0.45\linewidth}
        \centering
        \includegraphics[width=\linewidth]{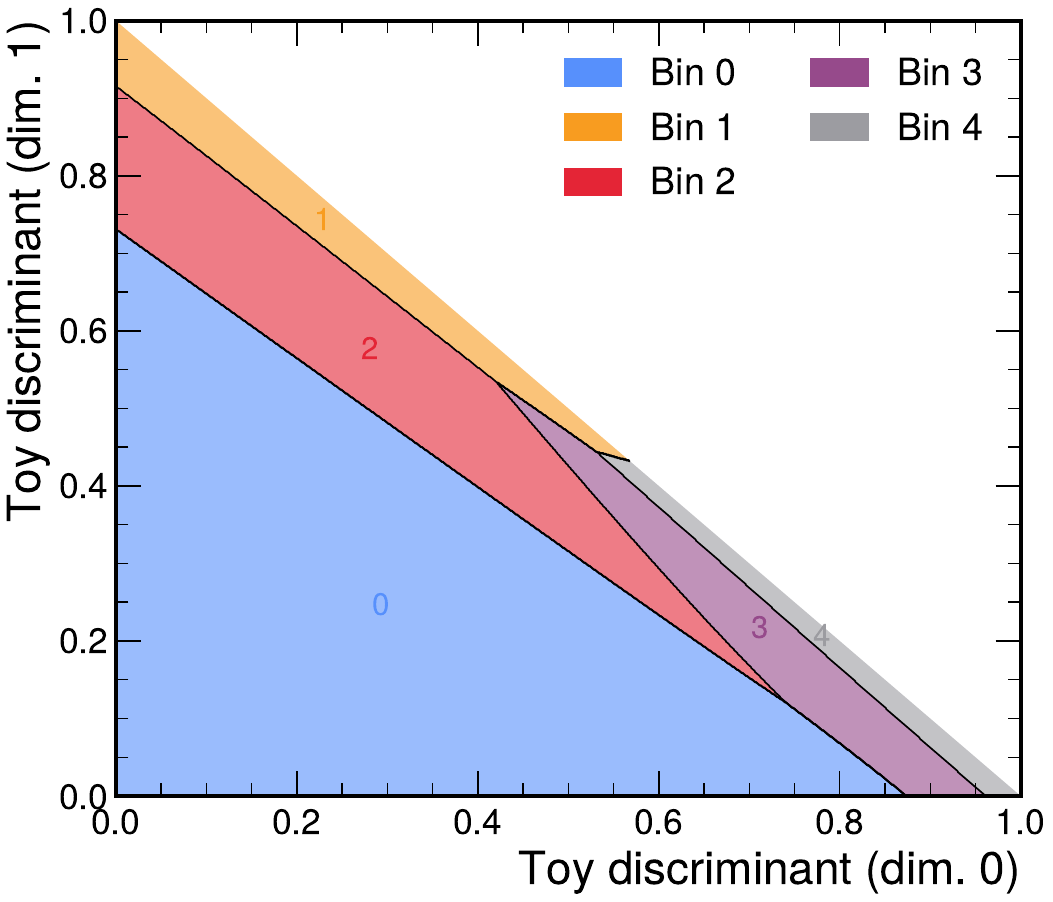}
        \caption{First scenario}
        \label{fig:Bin_boundaries_5_bins_scene_1}
    \end{subfigure}
    \hfill
    \begin{subfigure}[b]{0.45\linewidth}
        \centering
        \includegraphics[width=\linewidth]{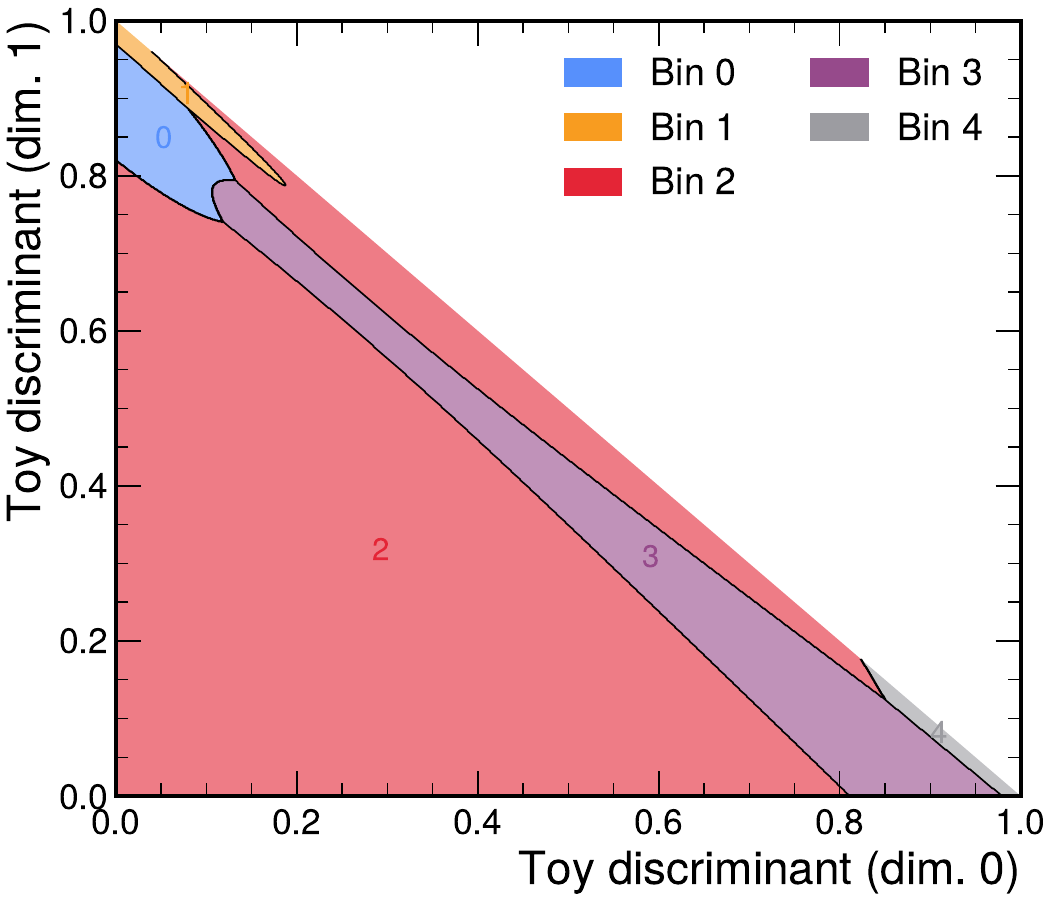}
        \caption{Second scenario}
        \label{fig:Bin_boundaries_5_bins_scene_2}
    \end{subfigure}
    \vspace{0.5cm}
    \begin{subfigure}[b]{0.45\linewidth}
        \centering
        \includegraphics[width=\linewidth]{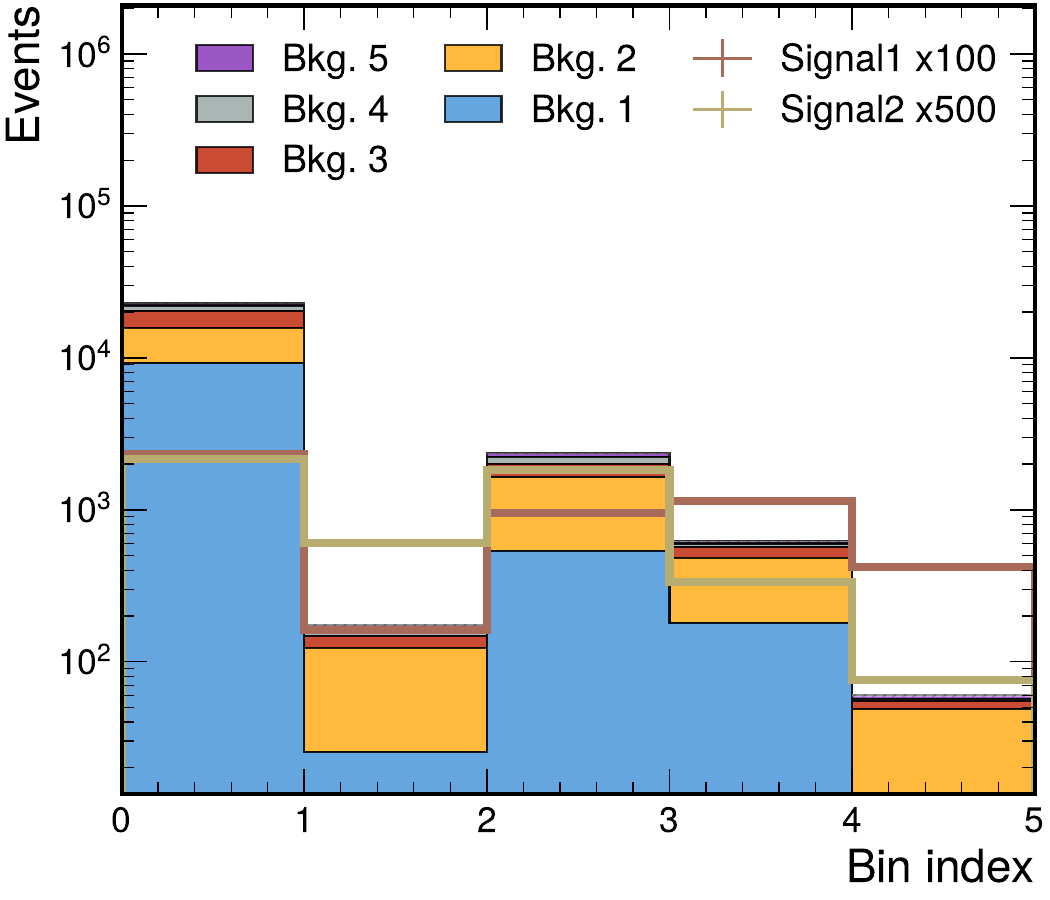}
        \caption{First scenario}
        \label{fig:gato_optimized_dist_5bins_log_scene1}
    \end{subfigure}
    \hfill
    \begin{subfigure}[b]{0.45\linewidth}
        \centering
        \includegraphics[width=\linewidth]{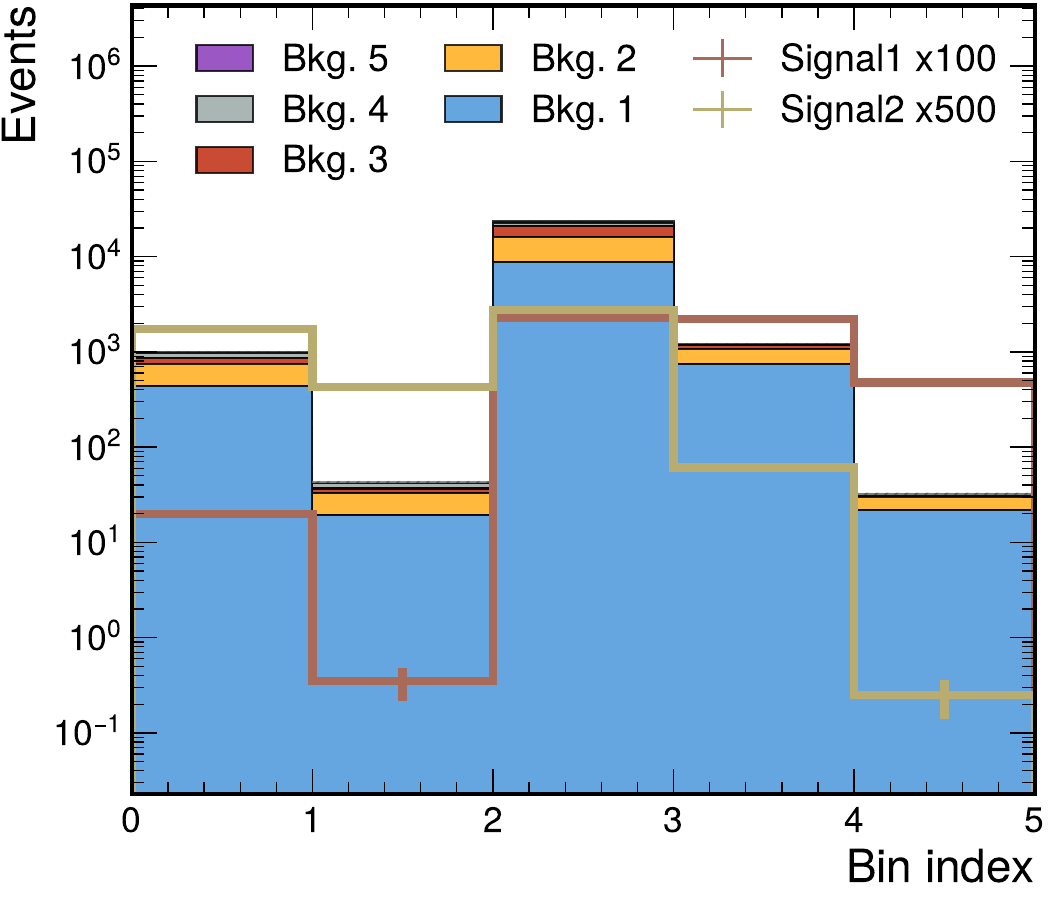}
        \caption{Second scenario}
        \label{fig:gato_optimized_dist_5bins_log_scene2}
    \end{subfigure}
    \caption{Top row: 68\% and 95\% density contours for the two signal processes and the background in the two-dimensional signal-score plane for the first (left) and second (right) scenario. Middle row: corresponding optimized five-bin boundaries in the same signal-score plane obtained with the GATO GMM approach. Bottom row: resulting five-bin event distributions for the two scenarios.}
    \label{fig:multi_bin_bound}
\end{figure}

Fig.~\ref{fig:multi_bin_bound} illustrates the two-dimensional contour plots of the two signal processes and the background in the signal-score plane (top row), together with the optimized bin boundaries (middle) and event distributions (bottom) for five bins, in both scenarios obtained from the GATO GMM approach.
The bin-boundary plots show how the solution adapts to the structure of the multi-dimensional score distributions, resulting in bin configurations aligned with regions of higher analysis sensitivity.
We observe that typically, a similar number of bins finds phase space regions that are sensitive to each of the signals, as expected from the symmetrically defined loss function.
The resulting expected event distributions in the five bins are shown in the plots in the bottom row.

\begin{figure}[ht]
    \centering
    \begin{subfigure}[b]{0.45\linewidth}
        \centering
        \includegraphics[width=\linewidth]{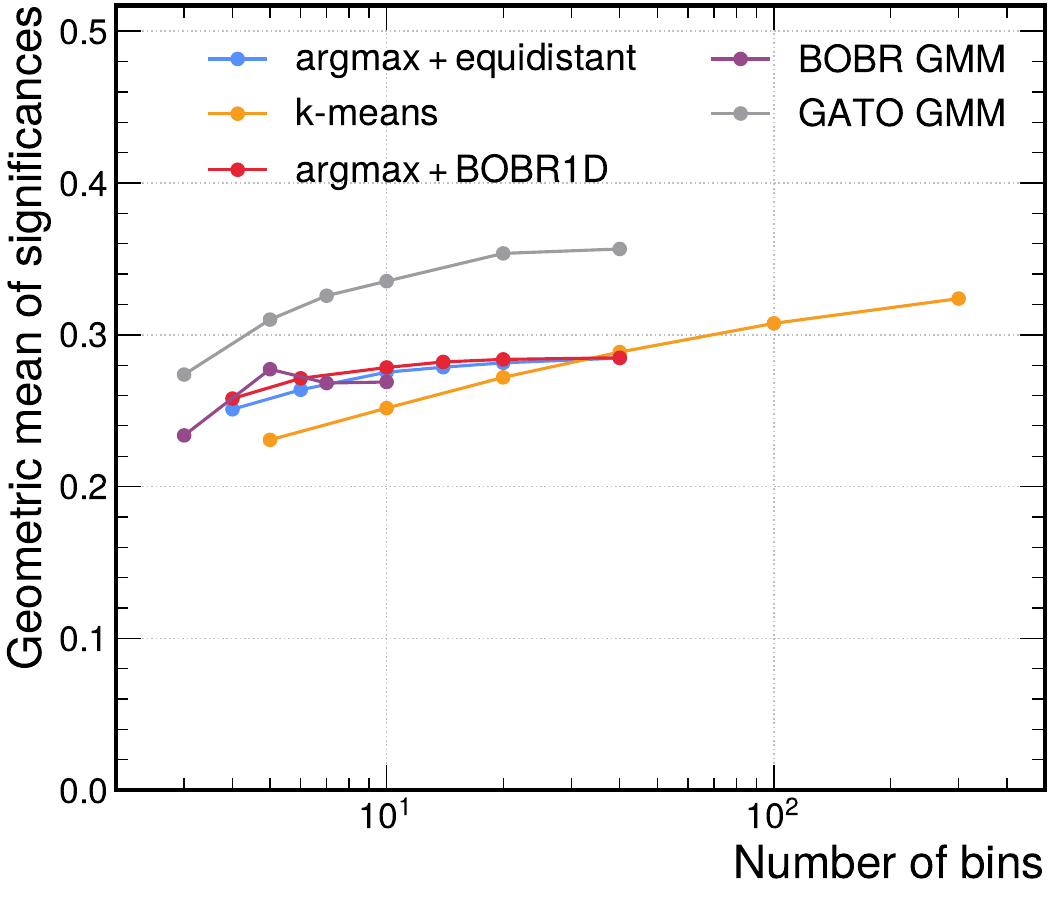}
        \caption{First scenario}
        \label{fig:3d_comp_case1}
    \end{subfigure}
    \hfill
    \begin{subfigure}[b]{0.45\linewidth}
        \centering
        \includegraphics[width=\linewidth]{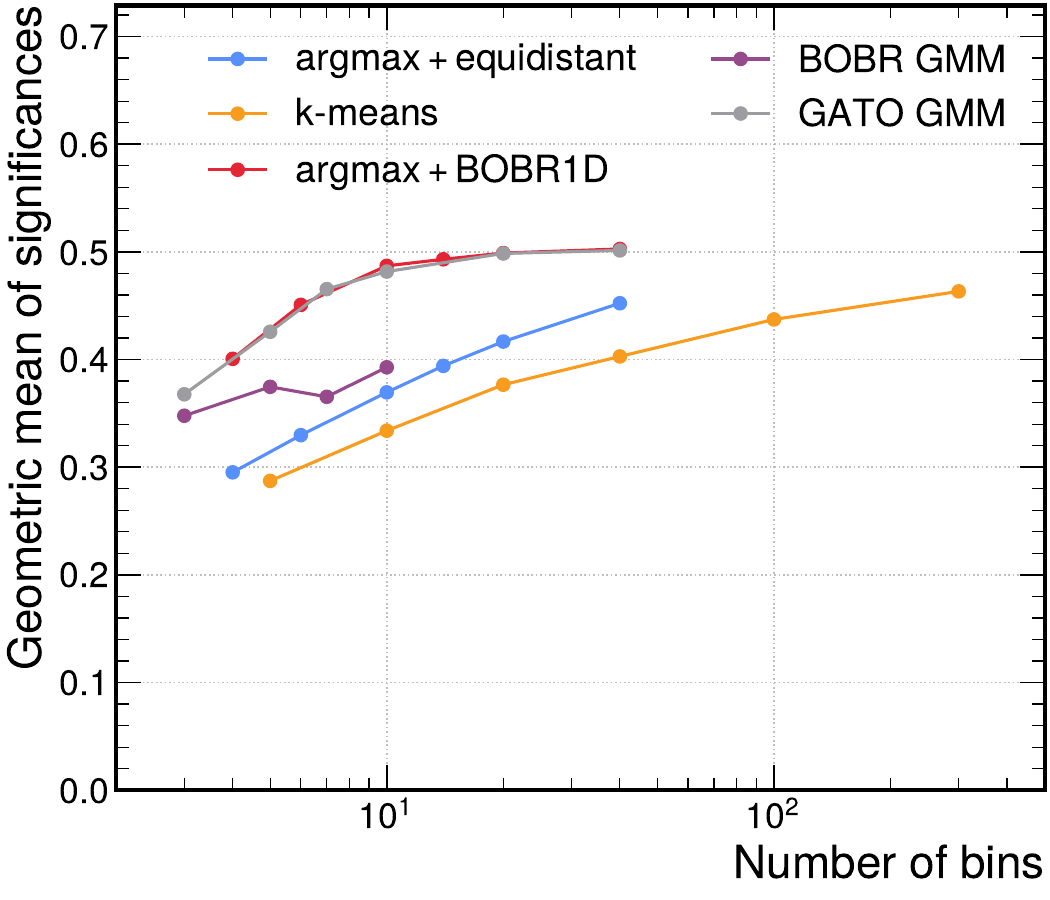}
        \caption{Second scenario}
        \label{fig:3d_comp_case2}
    \end{subfigure}
    \caption{Comparison of the geometric mean of the two signal significances of the different methods for the two representative scenarios of the three-dimensional toy discriminant. The $k$-means clustering approach is extended to much larger bin counts to illustrate its complementary scaling behavior, as its significance continues to improve.}
    \label{fig:3d_results}
\end{figure}

Fig.~\ref{fig:3d_results} compares the performance of the different methods in the two scenarios. 
In the first scenario, the GATO with GMM parameterization achieves the highest significance across all considered bin counts.
For a given bin number, it achieves approximately $20\,\%$ higher geometric mean of the two significances compared to all other methods.
This result highlights its ability to exploit the structure of the multi-dimensional score space and to define non-trivial bin boundaries.
The setups based on argmax classification with either equidistant or optimized binning in the two signal scores yield similar significances here and plateau already at small bin counts.
The $k$-means-clustered binning shows a stronger growth in significance as the number of bins increases, exceeding the argmax-based significance for bin counts of around 20.
However, as it is an unsupervised technique, it requires substantially larger bin counts than the GATO GMM approach.

In the second scenario, the GATO with GMM parameterization and the argmax+BOBR1D approach yield comparable significances.
The argmax+equidistant strategy also approaches the same significance of approximately 0.5, but only for substantially larger bin counts (around 150), which are beyond the range shown in Fig.~\ref{fig:3d_results}.
Here, the optimization problem factorizes to a good approximation into two separate optimization problems, one for each signal versus the background.
For both scenarios, in contrast to our findings in the one-dimensional optimization, the BOBR with GMM parameterization yields reasonable performance only for very low bin counts. As the parameter count increases, its performance suffers in comparison to gradient-based optimization with GATO.
This behavior can be attributed to the increasing dimensionality of the optimization problem, since each additional bin introduces several continuous parameters (means, covariances, and a mixture weight), which reduces the effectiveness of Bayesian optimization in moderately high-dimensional spaces.

\FloatBarrier

\subsubsection{Impact of penalties on low event counts or large statistical uncertainties}

In this subsection, we study the effect of the penalty terms introduced in Sec.~\ref{sec:binning_model} to regularize the binning optimization.
These penalties are designed to suppress bin configurations with very low expected background yields or large relative background uncertainties, which would otherwise lead to statistically unstable categories and are therefore disfavored in realistic HEP analyses.
A related idea has recently been explored in a fully differentiable analysis pipeline, where bins with insufficient Monte Carlo statistics are smoothly down-weighted or masked during training to stabilize the optimization~\cite{IceCube:2025sxb}.

To this end, we investigate the impact of these penalty terms in the second scenario, which is particularly sensitive to the presence of bins with low expected background yields and large statistical uncertainties.
The study is performed using the GATO GMM approach with a fixed configuration of five bins.

\begin{figure}[t]
    \begin{subfigure}[b]{0.45\linewidth}
        \centering
        \includegraphics[width=\linewidth]{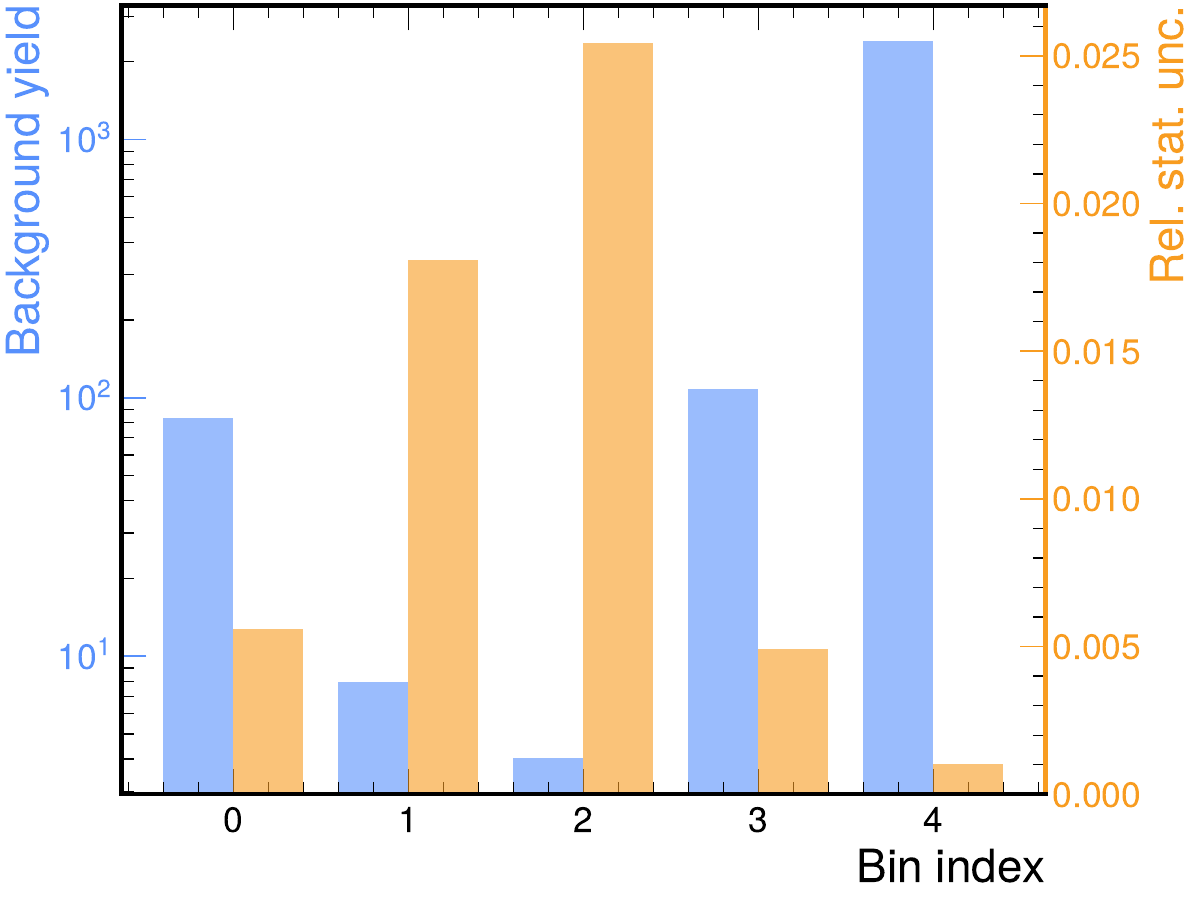}
        \caption{Low-background-yield penalty of one event and relative-uncertainty penalty of 50\%.}
        \label{fig:yield_vs_unc_5bins_log_redxsec}
    \end{subfigure}
    \hfill
    \begin{subfigure}[b]{0.45\linewidth}
        \centering
        \includegraphics[width=\linewidth]{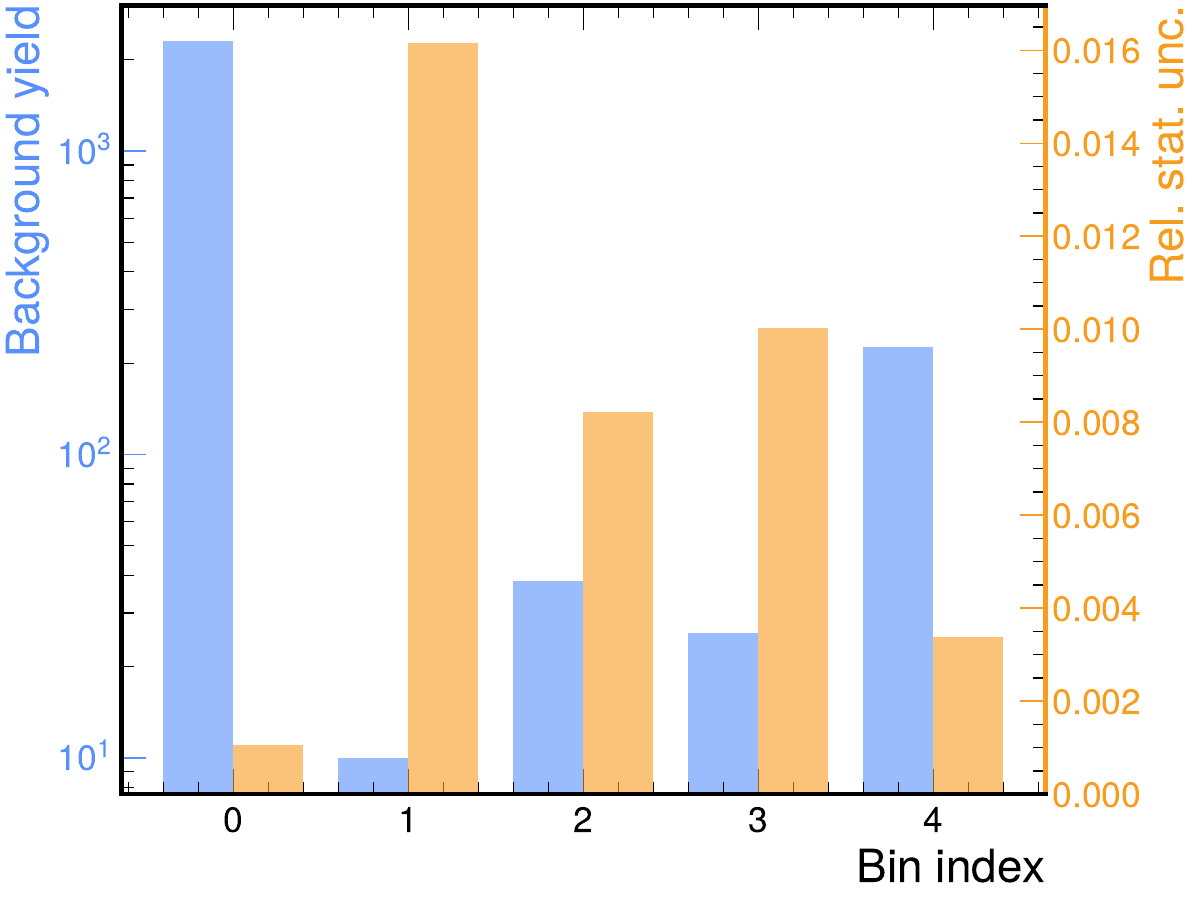}
        \caption{Low-background-yield penalty of ten events and relative-uncertainty penalty of 50\%.}
        \label{fig:yield_vs_unc_5bins_log_redxsec_with_penalty}
    \end{subfigure}

    \vspace{0.5cm}
    \centering
    \begin{subfigure}[b]{0.45\linewidth}
        \centering
        \includegraphics[width=\linewidth]{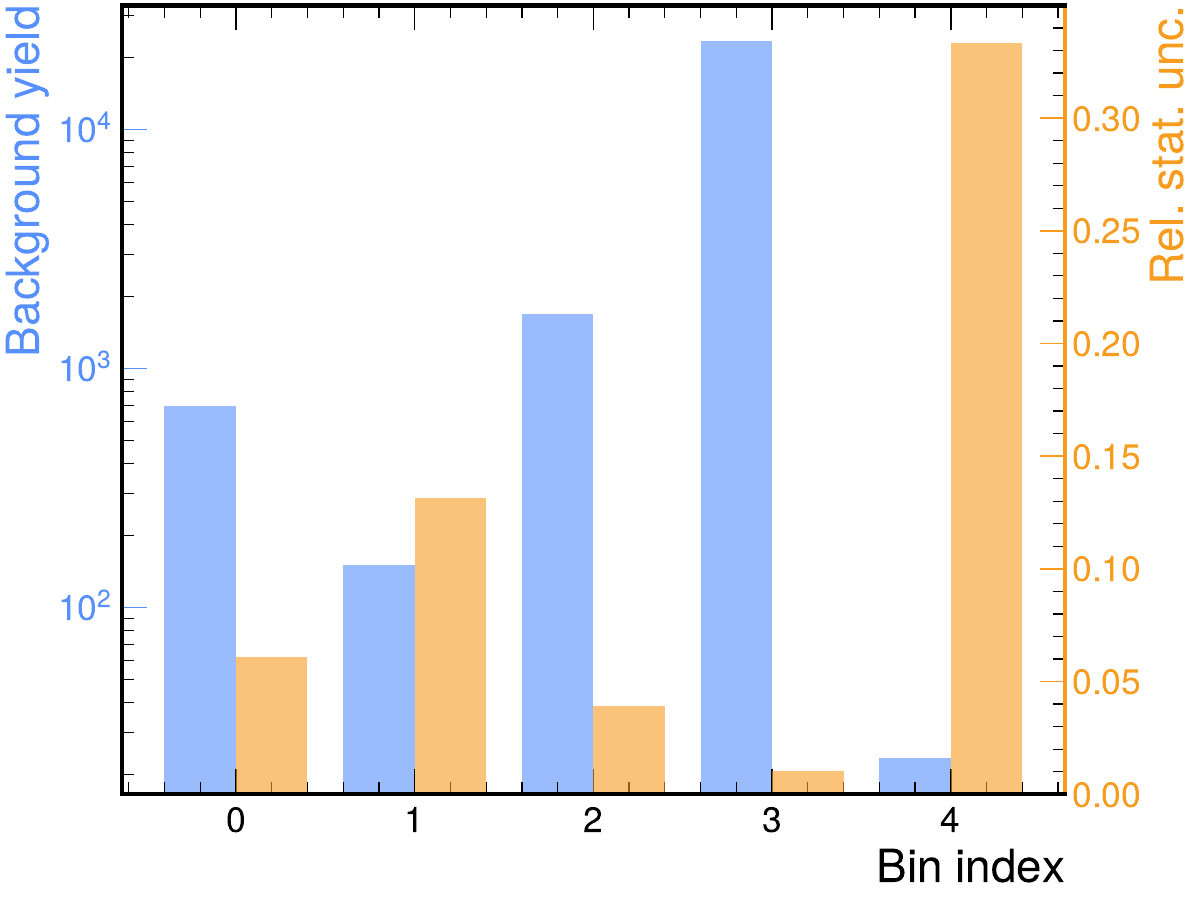}
        \caption{Low-background-yield penalty of one event and relative-uncertainty penalty of 50\%.}
        \label{fig:yield_vs_unc_5bins_log_red_stat}
    \end{subfigure}
    \hfill
    \begin{subfigure}[b]{0.45\linewidth}
        \centering
        \includegraphics[width=\linewidth]{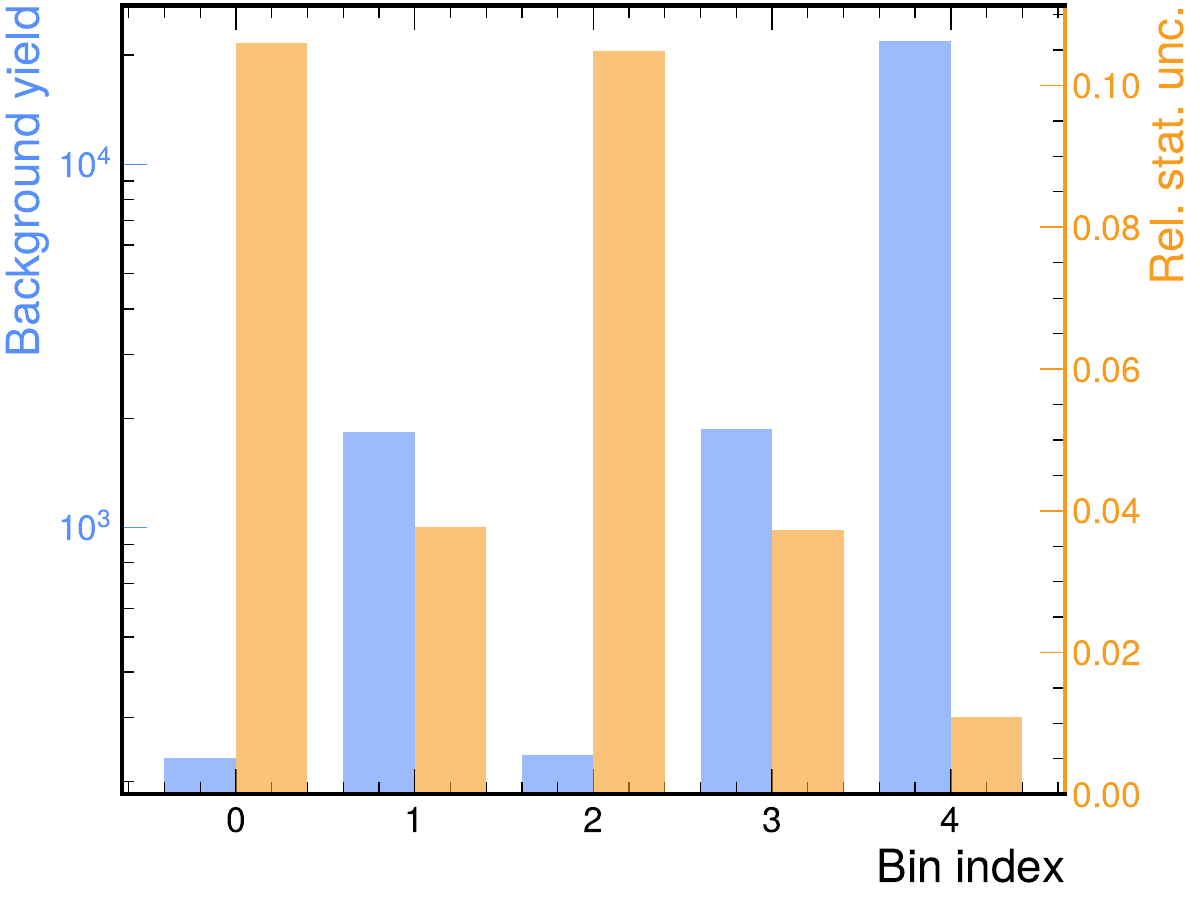}
        \caption{Low-background-yield penalty of one event and relative-uncertainty penalty of 10\%.}
        \label{fig:yield_vs_unc_5bins_linear_red_stat_with_penalty}
    \end{subfigure}

    \vspace{0.5cm}

    \caption{Expected background yield and relative background uncertainty per bin. Top row: background cross sections reduced by a factor of ten. Bottom row: total number of generated background events reduced to $10{,}000$.}
    \label{fig:penalty_plots}
\end{figure}

The effect of the low-background-yield penalty is first examined by reducing the
cross-sections of all background processes by a factor of ten, thereby introducing bins with lower expected background yields. 
To visualize the effect of this penalty, the minimum required number of expected background events per bin is then increased from one to ten, while keeping the relative background uncertainty threshold fixed at $50\%$.
In both cases, the corresponding penalty terms are included in the optimization objective with fixed weights, using \(\lambda = 0.02\) for the low-background-yield penalty and \(\lambda = 0.02\) for the relative-uncertainty penalty.
The results of this study are shown in the top row of Fig.~\ref{fig:penalty_plots}. 
Increasing the minimum background requirement guides the optimization toward configurations in which bin yields lie close to or above the chosen threshold.

The impact of the relative-background-uncertainty penalty is studied separately by reducing the total number of generated background events to $10{,}000$, increasing the statistical uncertainty per bin. 
In this case, the maximum allowed relative background uncertainty is tightened from $50\%$ to $10\%$, while maintaining the minimum background yield requirement at one event per bin. 
The corresponding penalty strengths are set to $\lambda = 0.02$ for the low-background-yield term and to $\lambda = 0.02$ and $\lambda = 30$ for the relative-uncertainty thresholds of $50\%$ and $10\%$, respectively.
The results of this study are shown in the bottom row of Fig.~\ref{fig:penalty_plots}. 
Tightening the upper bound on the relative background uncertainty leads to bin configurations in which the relative background uncertainty in the bins are below or around the threshold.

These results demonstrate that the penalty terms introduced in Section~\ref{sec:binning_model} effectively guide the optimization toward binning solutions with controlled background yields and statistical uncertainties, while retaining smoothness and flexibility in the optimization.

\FloatBarrier
\subsection{Performance study on the Fair Universe $\boldsymbol{H\rightarrow\tau\tau}$ dataset}
After studying the performance of the different methods in the toy datasets, we also study their behavior on a realistic public HEP dataset. 
For this purpose, we use the FAIR Universe HiggsML uncertainty challenge dataset~\cite{benato2025fairuniversehiggsmluncertainty}, which is based on simulated proton--proton collision events representative of data collected by the ATLAS experiment at the Large Hadron Collider. 
The dataset consists of the $H\rightarrow\tau\tau$ signal process and three background processes with contributions from $Z\rightarrow\tau\tau$, $t\bar{t}$, and diboson ($VV$) production, with each event represented by 28 features. 
Further details of the dataset construction can be found in Ref.~\cite{benato2025fairuniversehiggsmluncertainty}.

In this study, we train a four-class softmax classifier using the signal and the three background processes as separate classes.
Details of the classifier architecture and training setup are provided in Appendix~\ref{app:training}.
The resulting classifier scores are then used as input to the binning optimization methods, with the objective of maximizing the expected $H\rightarrow\tau\tau$ signal significance.
The optimization is performed on an independent set of 5,000,000 events not used for training or validation, with event weights rescaled to the full dataset normalization when computing the expected significance.

\begin{figure}[h!]
    \centering
    \includegraphics[width=0.45\linewidth]{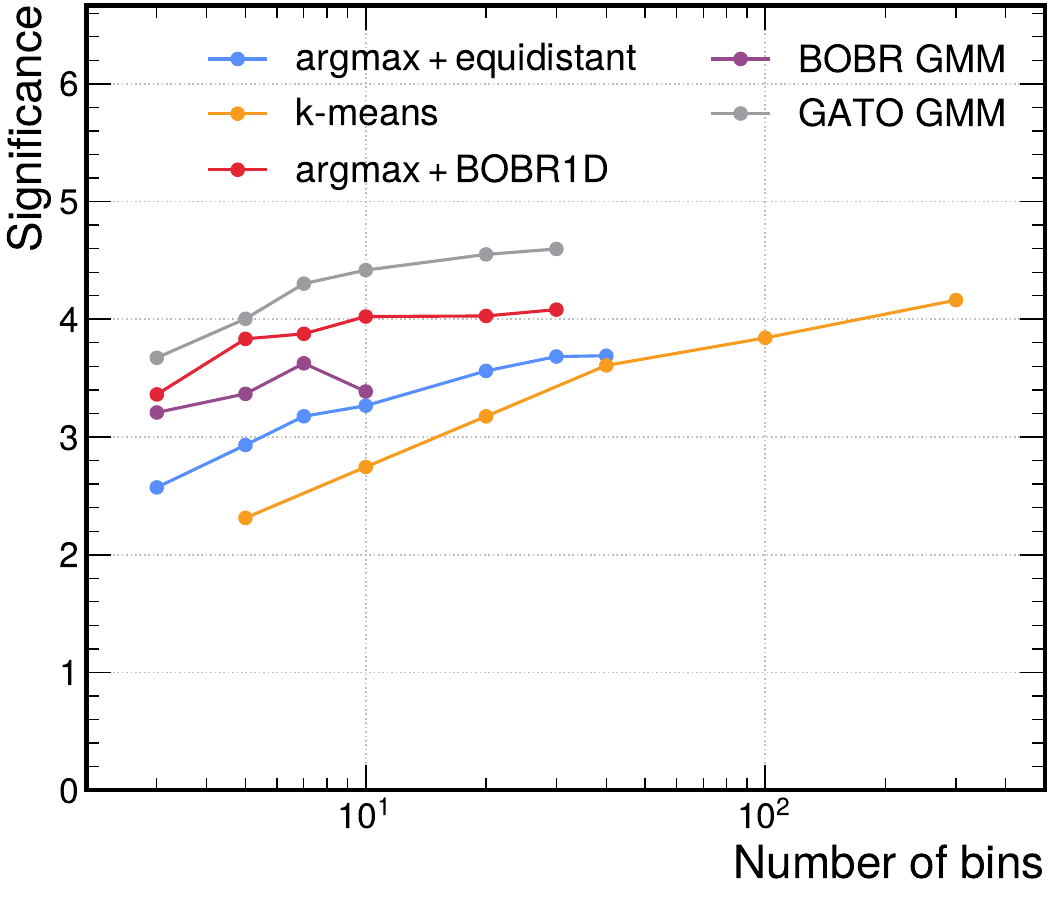}
    \caption{Comparison of the expected $H\rightarrow\tau\tau$ significance with different binning methods on the FAIR Universe HiggsML uncertainty dataset.
    }
    \label{fig:3d_comp_fair}
\end{figure}

In Figure~\ref{fig:3d_comp_fair}, the expected $H\rightarrow\tau\tau$ significances obtained with the different binning methods are compared.
For argmax-based approaches, events are first assigned to the class with the highest classifier score, after which only the $H\rightarrow\tau\tau$ signal category is optimized in one dimension. 
For the GMM-based parameterization, the softmax normalization constrains the four multi-classifier scores to sum to unity, so only three scores are independent and one score can be omitted without loss of information.
In this study, the $VV$ score is omitted, and the remaining three scores are used as inputs to the GMM-based optimization.
The GATO models with GMM parameterization achieve the highest significance for all considered bin counts, improving over argmax+BOBR1D by approximately \(15\%\).
BOBR with the GMM parameterization performs better than argmax+equidistant binning, while the $k$-means clustering approach shows a scaling behavior with the number of bins similar to that observed in the toy setups.

\FloatBarrier
\section{Conclusions}

We studied the optimization of bin boundaries with respect to signal significance in the context of high-energy physics analyses that perform inference with binned likelihoods. The proposed techniques can also be applied in the case of unbinned likelihoods that use different event categories, where the category boundaries are then optimized.
Since binning compresses the underlying event-level information into counts per bin, the choice of how to categorize within a discriminant space directly impacts the sensitivity to the chosen figure of merit.

We introduced a categorization approach, where events are assigned to the component of a weighted Gaussian mixture that yields the largest probability density at the event’s discriminant value, applicable to both one- and multi-dimensional discriminants.
This approach allows bin boundaries with flexible shapes to be learned directly in the full discriminant space, without the need to use one-dimensional projections, and can be optimized for signal significance or any function of the binned signal and background yields.
On top of this common binning model we provide two numerical optimization strategies: a gradient-descent based approach (GATO), relying on a differentiable approximation of the assignment achieved with a temperature-scaled softmax, and a Bayesian optimization approach using a tree-structured Parzen estimator (BOBR).

The methods were benchmarked on toy setups constructed from three-dimensional Gaussian processes with one or two signals and five backgrounds, from which we built two analytic likelihood-ratio-based discriminants: a one-dimensional binary discriminant separating one signal from the combined background, and a three-class softmax score that separates two distinct signals from the background.
In the one-dimensional setup, when optimizing the bin positions directly, GATO and BOBR yield similarly sensitive categorization schemes, and both clearly outperform a fixed equidistant binning for a given number of bins. 
In the three-class setup, the multi-dimensional GMM-based binning defines categories with non-trivial boundary shapes in the signal-score plane.
As the number of free parameters increases, the differentiable GATO approach shows its advantages over Bayesian optimization with BOBR.
We compared the achieved signal sensitivity against several baseline scenarios, including argmax classification followed by one-dimensional optimizations in each signal-component score, and $k$-means clustering of the score itself.
The potential gain from the GMM-based optimization is governed by the structure of the multi-dimensional score distributions. In the first toy scenario, where the two signals and the background substantially overlap in the signal-score plane, the multi-dimensional optimization provides clear significance improvements over argmax-based projections with subsequent one-dimensional optimization, and can reach a given significance with substantially fewer bins than the baseline methods. In the second scenario, where the signal-enriched regions are more localized, the projected approaches with one-dimensional optimization become competitive.

We further tested the methods on the FAIR Universe $H\rightarrow\tau\tau$ dataset, a public benchmark dataset, using the output scores of a four-class classifier trained to distinguish the signal from $Z\rightarrow\tau\tau$, $t\bar{t}$, and diboson production.
This extends the comparison beyond the toy discriminants to classifier outputs obtained from a realistic multi-class setup. 
Among the tested methods, the GATO with GMM parameterization provides the highest expected significance across all the considered bin counts, improving over argmax with subsequent one-dimensional binning optimization by approximately \(15\%\).

Beyond the choice of figure of merit, we demonstrated that the framework can incorporate additional penalty terms to encode practical analysis constraints.
In particular, penalizing bins with too low expected event yields or too large relative uncertainties in the process predictions (Monte Carlo statistical uncertainties) effectively steers the optimization away from impractical configurations and toward robust, analysis-friendly binning schemes.
Overall, the proposed GMM-based binning model, together with its gradient-based (GATO) and Bayesian (BOBR) optimization strategies, provides a flexible and analysis-aware alternative to hand-crafted binning choices.

Implemented and released as the lightweight \texttt{gato-hep} and \texttt{bobr-hep} Python packages, these methods are ready to be integrated into existing analyses with one- or multi-dimensional discriminants.
A natural next step for further research may be to embed the GMM-based categorization with GATO into a fully differentiable pipeline that jointly optimizes the classifier and the binning using likelihood-based objectives, such as the expected uncertainty on the parameter of interest.
Keeping the classifier, binning model, and statistical inference as explicit modules rather than a single monolithic block may offer advantages in interpretability and validation, while still enabling end-to-end optimization of the overall analysis sensitivity.

\subsubsection*{Acknowledgements}

This research was supported by the Deutsche Forschungsgemeinschaft (DFG) under grants 400140256 - GRK~2497 (The physics of the heaviest particles at the LHC, all authors) and 686709 - ER~866/1-1 (Heisenberg Programme, JE),
by the German Federal Ministry of Research, Technology and Space (BMFTR) via project 05H2024 under grant number 05H24PA4 (FM), and by the Studienstiftung des deutschen Volkes (FM).
The authors thank Svenja Diekmann, Martin Erdmann, Benjamin Fischer and Felix Zinn for useful discussions and for feedback on the $k$-means clustering approach.

\appendix
\section{Data generation details}
\label{app:dataset}
This appendix documents the construction of the toy discriminants used in this study.

Each physics process $i$ (signals and backgrounds) is represented by a
three-dimensional probability density.  For an event belonging to
process~$i$, the feature vector \(X_i \in \mathbb{R}^3\) is drawn from a
multivariate normal distribution
\begin{equation*}
X_i \sim \mathcal{N}(\mu_i,\Sigma),
\end{equation*}
with mean vector \(\mu_i\) and a common covariance matrix
\begin{equation*}
\Sigma = I_{3\times 3} + 0.2\,(\mathbf{1}_{3\times 3} - I_{3\times 3}).
\end{equation*}
Here, $I_{3\times 3}$ denotes the $3\times 3$ identity matrix and $\mathbf{1}_{3\times 3}$ is the $3\times 3$ matrix of ones.
To avoid unrealistically sharp separation between processes in this simplified toy setup, we apply a multiplicative Gaussian perturbation $\mathcal{N}(1,\,0.1^{2})$ to the sampled observables. 

A toy cross section $\sigma_i$ is assigned to each process.
Then, a number of events \(N_i\) is drawn for each process, and we assign weights
\begin{equation*}
w_i = \frac{\sigma_i\,\mathcal{L}}{N_i},
\end{equation*}
with a toy luminosity of $\mathcal{L}=100\,\mathrm{fb}^{-1}$, reflecting a fully efficient selection.
In both the 1D and 3D examples, the total number of background events is partitioned among the five background processes in proportion to their cross sections.
The means of the Gaussians and the associated production cross sections for all processes are listed in Table~\ref{tab:gauss_params}.
Here, we include two scenarios considered for the three-class study.
In the second scenario, the signal means and selected background means are shifted relative to the first scenario, changing how the generated signal and background events populate the resulting classifier-score space.

The total background likelihood for an event is hence given by 
\begin{equation*}
      p_{\mathrm{bkg.}}(X)=\sum_{i \in \{\mathrm{bkg.\,1},\dots,\mathrm{bkg.\,5}\}} \frac{\sigma_i}{\sum_{j\in \{\mathrm{bkg.\,1},\dots,\mathrm{bkg.\,5}\}}\sigma_j}\;\mathcal{N}(X\mid\mu_i,\Sigma)\,.
\end{equation*}
This background density is therefore a mixture of the five background components with mixture weights proportional to their cross sections.
We then build the likelihood‐ratio discriminant $D$, which is normalized to range from zero to one as
\begin{equation*}
  L(X)=\frac{p_{\rm sig.\,1}(X)}{p_{\mathrm{bkg.}}(X)}\,,\quad
  D(X)=\frac{L(X)}{1+L(X)}\in[0,1].
\end{equation*}

For the three-class example, we also include Signal 2.
We denote densities by $p_{\mathrm{sig.\,1}}(X)$, $p_{\mathrm{sig.\,2}}(X)$, and the background mixture 
density $p_{\mathrm{bkg.}}(X)$ defined above. We then construct a softmax-like multi-class output
\begin{equation*}
S_k(X)=
\frac{p_k(X)}{
p_{\mathrm{sig.\,1}}(X) + p_{\mathrm{sig.\,2}}(X) + p_{\mathrm{bkg.}}(X)
},
\end{equation*}
where the index $k$ denotes either Signal 1, Signal 2 or the total background.

\begin{table}[ht]
  \centering
  \sisetup{
    table-number-alignment = center,
    table-figures-integer  = 1,
    table-figures-decimal  = 1,
    table-space-text-post  = \,,
  }
\caption{Mean components $(\mu_x,\mu_y,\mu_z)$ and toy production cross-sections $\sigma$ for all Gaussian processes in the two toy setups.
Scenario 1 and Scenario 2 correspond to two alternative choices of Gaussian mean vectors for the signal and background processes, leading to different multidimensional classifier-score configurations.
All components share the common covariance $\Sigma = I_{3\times 3} + 0.2\,(\mathbf{1}_{3\times 3} - I_{3\times 3})$, where $I_{3\times 3}$ is the $3\times 3$ identity matrix and $\mathbf{1}_{3\times 3}$ the $3\times 3$ matrix of ones.
Entries marked with a hyphen in Scenario~2 are identical to the corresponding Scenario~1 values.}
  \label{tab:gauss_params}

  \begin{tabular}{l
                  S[table-format=1.1]
                  S[table-format=1.1]
                  S[table-format=1.1]
                  S[table-format=1.1]
                  S[table-format=1.1]
                  S[table-format=1.1]
                  S[table-format=3.1]}
    \toprule
    & \multicolumn{3}{c}{Scenario 1} & \multicolumn{3}{c}{Scenario 2} & \\
    \cmidrule(lr){2-4}\cmidrule(lr){5-7}
    {Process} & {$\mu_x^{(1)}$} & {$\mu_y^{(1)}$} & {$\mu_z^{(1)}$}
              & {$\mu_x^{(2)}$} & {$\mu_y^{(2)}$} & {$\mu_z^{(2)}$}
              & {$\sigma\,[\mathrm{pb}]$} \\
    \midrule
    Signal 1      &  0.4 & -0.4 &  1.0 &  0.9 & -0.9 &  1.0 &   0.5 \\
    Signal 2      & -0.4 &  0.4 &  1.0 & -0.9 &  0.9 &  1.0 &   0.1 \\
    Background 1  &  0.0 &  0.0 & -0.5 &  0.2 &  0.2 &  0.5 & 100.0 \\
    Background 2  & -0.2 &  0.0 &  0.0 &
      \multicolumn{1}{c}{--} &
      \multicolumn{1}{c}{--} &
      \multicolumn{1}{c}{--} &  80.0 \\
    Background 3  &  0.1 &  0.4 & -0.3 &
      \multicolumn{1}{c}{--} &
      \multicolumn{1}{c}{--} &
      \multicolumn{1}{c}{--} &  50.0 \\
    Background 4  & -0.1 &  0.6 & -0.2 &
      \multicolumn{1}{c}{--} &
      \multicolumn{1}{c}{--} &
      \multicolumn{1}{c}{--} &  20.0 \\
    Background 5  & -0.2 &  0.1 & -0.1 &
      \multicolumn{1}{c}{--} &
      \multicolumn{1}{c}{--} &
      \multicolumn{1}{c}{--} &  10.0 \\
    \bottomrule
  \end{tabular}
\end{table}

\section{Multi-class training setup}
\label{app:training}
From the available FAIR Universe $H\rightarrow\tau\tau$ public dataset, we construct a stratified subset of 5,000,000 events, where the number of sampled events from each process is chosen proportionally to the number of generated events in the original sample.
This subset is divided into training and validation samples using a 70:30 ratio. 
During training and validation, the event weights are normalized separately for each process such that the total weight of each process is unity.
These normalized weights are used to compute a weighted categorical cross-entropy loss.
The classifier is trained using the 28 available input features, which are standardized using the mean and standard deviation computed from the training sample.

We use a fully connected neural network implemented in TensorFlow~\cite{abadi2016tensorflowlargescalemachinelearning}, with three hidden layers of sizes 128, 64 and 32, respectively. 
Each hidden layer uses a ReLU activation, followed by dropout~\cite{hinton2012improvingneuralnetworkspreventing,JMLR:v15:srivastava14a} with a rate of 0.1. 
Weight regularization is implemented as $L_2$ regularization with coefficient $10^{-4}$.
The output layer consists of four nodes, corresponding to the four physics processes, and uses a softmax activation. 
The network is trained using the Adam~\cite{kingma2017adammethodstochasticoptimization} optimizer with an initial learning rate of $10^{-3}$, a batch size of 512, and a maximum of 200 epochs. 
Early stopping is applied if the validation loss does not improve for 30 consecutive epochs.

\clearpage

\phantomsection
\addcontentsline{toc}{section}{References}
\bibliographystyle{JHEP_jls}
\bibliography{main.bib}

\end{document}